\documentclass[sigconf]{acmart}

\copyrightyear{2026}
\acmYear{2026}
\setcopyright{cc}
\setcctype{by}
\acmConference[CIKM '26]{Proceedings of the 35th ACM International Conference on Information and Knowledge Management}{November 07--11, 2026}{Rome, Italy}
\acmBooktitle{Proceedings of the 35th ACM International Conference on Information and Knowledge Management (CIKM '26), November 07--11, 2026, Rome, Italy}
\acmISBN{979-8-4007-2539-5/2026/11}
\acmDOI{10.1145/3799682.3840701}

\AtBeginDocument{%
  }

\usepackage{amsmath}
\usepackage{amsfonts}
\usepackage{amsthm}
\usepackage{multirow}
\usepackage{makecell}
\usepackage[most]{tcolorbox}
\usepackage{enumitem}

\definecolor{lightgreen}{RGB}{144,238,144}
\definecolor{target}{HTML}{32CD32}
\definecolor{unpredicted}{HTML}{F28C8C}
\definecolor{candidate}{HTML}{87CEFA}
\definecolor{history}{HTML}{D3D3D3}

\makeatletter
\@ifundefined{theorem}{}{}
\@ifundefined{lemma}{}{}
\@ifundefined{proposition}{}{}
\makeatother

\title[Recommender System as Slow and Fast Thinkers]{Recommender System as Slow and Fast Thinkers}

\author{Zichen Yuan}
\affiliation{%
  \institution{City University of Hong Kong}
  \city{Hong Kong}
  \country{Hong Kong}
}
\email{yzc66633@gmail.com}

\author{Xiaoxuan Dong}
\affiliation{%
  \institution{University of Electronic Science and Technology of China}
  \city{Chengdu}
  \country{China}
}
\email{202522010524@std.uestc.edu.cn}

\author{Linkun Dai}
\affiliation{%
  \institution{Shanghai Jiao Tong University}
  \city{Shanghai}
  \country{China}
}
\email{dailinkun@sjtu.edu.cn}

\author{Jinwei Yang}
\affiliation{%
  \institution{University of Electronic Science and Technology of China}
  \city{Chengdu}
  \country{China}
}
\email{202522010132@std.uestc.edu.cn}

\author{Jining Luan}
\affiliation{%
  \institution{University of Electronic Science and Technology of China}
  \city{Chengdu}
  \country{China}
}
\email{202521081036@std.uestc.edu.cn}

\author{Dexu Yu}
\affiliation{%
  \institution{Fenz.AI}
  \city{Palo Alto}
  \country{United States}
}
\email{yu.dex@northeastern.edu}

\author{Chunxiao Li}
\affiliation{%
  \institution{University of Science and Technology of China}
  \city{Hefei}
  \country{China}
}
\email{chunxiao.li@ustc.edu.cn}

\author{Joemon M. Jose}
\affiliation{%
  \institution{GAIR-Lab - School of Computing Science,}
  \institution{University of Glasgow}
  \city{Glasgow}
  \country{United Kingdom}
}
\email{joemon.jose@glasgow.ac.uk}

\author{Youhua Li}
\affiliation{%
  \institution{City University of Hong Kong}
  \city{Hong Kong}
  \country{Hong Kong}
}
\email{youhuali2-c@my.cityu.edu.hk}

\author{Hanwen Du}
\authornote{Corresponding authors.}
\affiliation{%
  \institution{The Ohio State University}
  \city{Columbus}
  \country{United States}
}
\email{du.1128@osu.edu}

\author{Junchen Fu}
\authornotemark[1]
\affiliation{%
  \institution{GAIR-Lab - School of Computing Science,}
  \institution{University of Glasgow}
  \city{Glasgow}
  \country{United Kingdom}
}
\email{j.fu.3@research.gla.ac.uk}
\renewcommand{\shortauthors}{Zichen Yuan et al.}

\begin{abstract}
Sequential recommendation models are foundational to modern personalized services, yet their effectiveness varies substantially across heterogeneous user environments. In particular, static one-pass recommenders often perform well on common behavior patterns but degrade on operationally challenging user groups, such as users with longer histories or less mainstream item profiles. To address this limitation, we propose \textsc{DS-Frame}, an adaptive fast--slow inference framework for sequential recommendation. \textsc{DS-Frame} combines a Fast System for efficient routine prediction, a Slow System for iterative latent refinement, and a learned selector that routes each sample under a controllable computation budget. Experiments on five real-world datasets show that \textsc{DS-Frame} consistently improves representative sequential recommendation backbones, with larger gains on challenging groups and effective accuracy--efficiency trade-offs. This highlights the potential of adaptive inference for more efficient and robust recommendation. Code is available at \href{https://github.com/ZichenYuan233/Recommender-System-as-Slow-and-Fast-Thinkers}{this link}.
\end{abstract}



\begin{document}

\maketitle
\begin{figure}
    \centering
    \includegraphics[
        width=1.08\linewidth,
        trim=90 100 90 5,
        clip
    ]{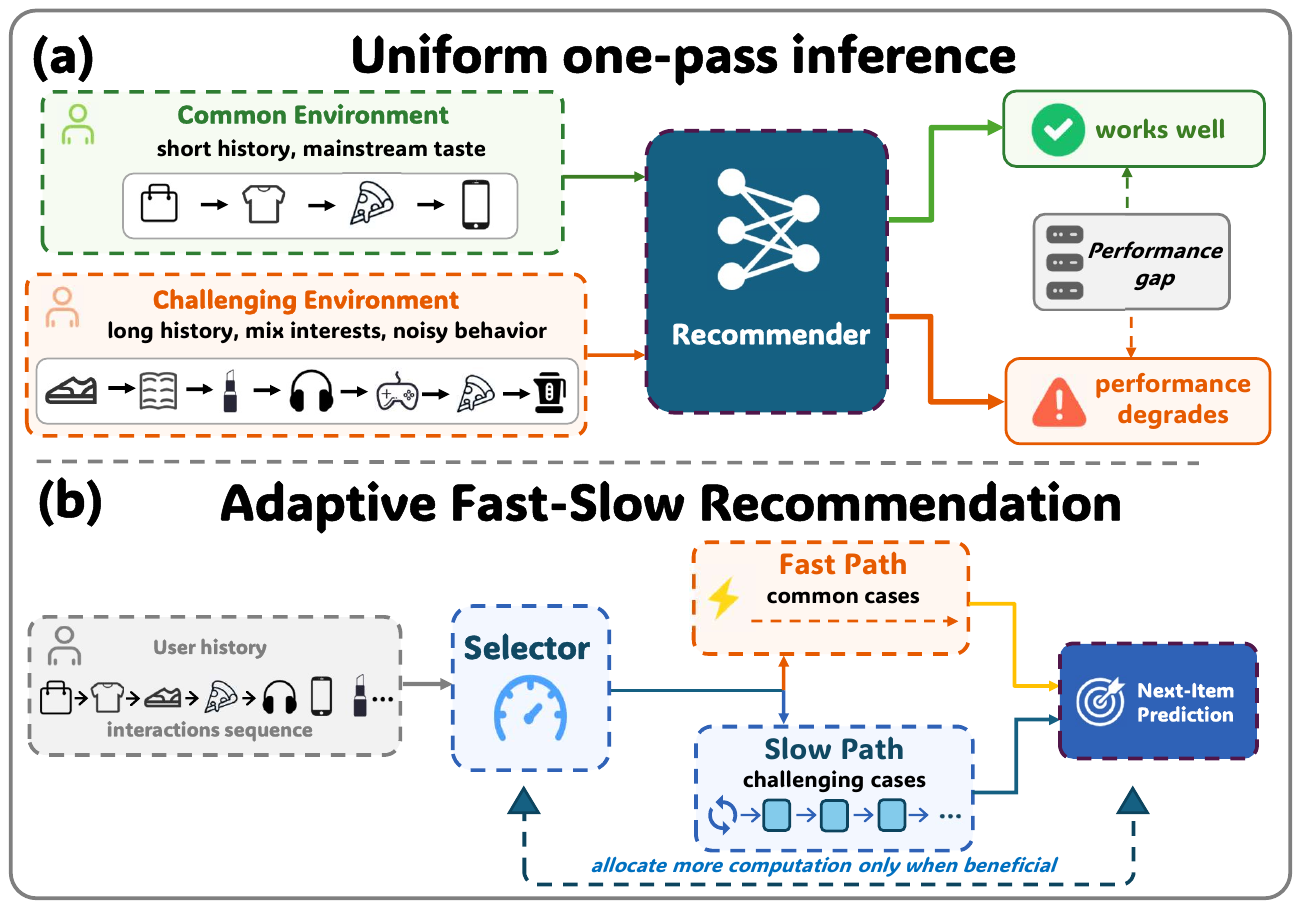}
   \caption{Motivation of \textsc{DS-Frame}. Static one-pass recommenders show uneven performance across heterogeneous user groups, motivating adaptive fast--slow inference that allocates more computation to harder cases.}
    \Description{The figure provides a high-level illustration of the dual-system recommendation idea, where a fast pathway handles common cases and a slow pathway provides additional reasoning for challenging cases.}
    \label{fig:intro_model}
\end{figure}
\section{Introduction}
Sequential recommendation aims to predict a user's next interaction from their historical behavior sequence, and has become a core technique behind modern personalized services such as e-commerce, media streaming, and online content platforms. Early neural methods model user dynamics with recurrent networks~\cite{hidasi2015session,quadrana2017personalizing}, while later convolutional, graph-based, and self-attentive models capture higher-order transition patterns and long-range dependencies~\cite{tang2018personalized,wu2019session,chang2021sequential,kang2018self,sun2019bert4rec}.

Despite their architectural differences, existing sequential recommenders largely follow the same static inference principle: once trained, every user sequence is processed by the same computational graph and receives an identical amount of computation at test time. This paradigm can be effective for routine cases, where short histories and mainstream behavior patterns allow the next item to be inferred from local transition signals. However, real users differ substantially in history length, interest consistency, and preference mainstreamness~\cite{hu2022memory,Ungruh2025monolith,dang2026tail}. For more complex cases, a single forward pass may fail to resolve competing interests, overlook weak but important signals scattered across long histories, or produce an under-refined representation of user intent~\cite{ma2020disentangled,chen2022intent,wang2024spark}. The key issue is therefore not merely that some users are harder to predict, but that static inference allocates the same computational budget regardless of whether additional computation would be useful.

We empirically verify this limitation by segmenting users according to interaction sequence length and historical item popularity. A single model trained on the full dataset performs well for users with shorter histories and mainstream item profiles, which we refer to as the ``common environment''. In contrast, its performance degrades on operationally harder groups, such as users with longer histories or less mainstream item profiles, which we refer to as the ``challenging environment''. These results suggest that the standard one-size-fits-all inference strategy is mismatched with the heterogeneous nature of real recommendation environments.

This observation raises an interesting research question: \emph{Can existing sequential recommenders be equipped with an adaptive inference mechanism that allocates additional computation only when it is likely to improve prediction?} To answer this question, we draw inspiration from Dual Process Theory (DPT) in cognitive science~\cite{kahneman2003maps}, which distinguishes between a fast, intuitive system for routine tasks and a slower, deliberative system for complex ones. This perspective provides a natural computation-allocation principle for recommendation: common cases should be handled by a lightweight fast path, while challenging cases should selectively invoke a more expressive slow path.

Based on this insight, we propose \textsc{DS-Frame}, a plug-and-play fast--slow inference framework for sequential recommendation. \textsc{DS-Frame} uses an existing backbone recommender as the Fast System for efficient one-pass prediction, and attaches a Slow System that performs iterative latent refinement for samples requiring deeper inference. Crucially, \textsc{DS-Frame} does not apply slow reasoning uniformly. Instead, it trains a lightweight selector with oracle-guided supervision and budget regularization to estimate the sample-level benefit of slow refinement. At inference time, the selector routes each user sequence to either the fast or slow path, enabling an adaptive accuracy--efficiency trade-off under a constrained budget.

Our contributions are as follows:
\begin{itemize}[leftmargin=*, topsep=0pt]
    \item \textbf{User-level Heterogeneity Analysis.}
    We empirically characterize performance gaps of static sequential recommenders across operational user groups defined by interaction length and historical item popularity.

    \item \textbf{Adaptive Fast--Slow Inference.}
    We formulate sequential recommendation inference as a sample-level computation allocation problem and propose \textsc{DS-Frame}, a plug-and-play framework that combines a Fast System, an iterative latent-refinement Slow System, and a selector under a target computation budget.

    \item \textbf{Routing and Generality Validation.}
    We instantiate \textsc{DS-Frame} on representative sequential recommendation backbones and validate its overall performance, group-wise gains, routing behavior, and accuracy--efficiency trade-offs.
\end{itemize}

\begin{figure}
    \centering
    \includegraphics[
        width=\linewidth,
        trim={0cm 0.2cm 0.1cm 2.2cm},
        clip
    ]{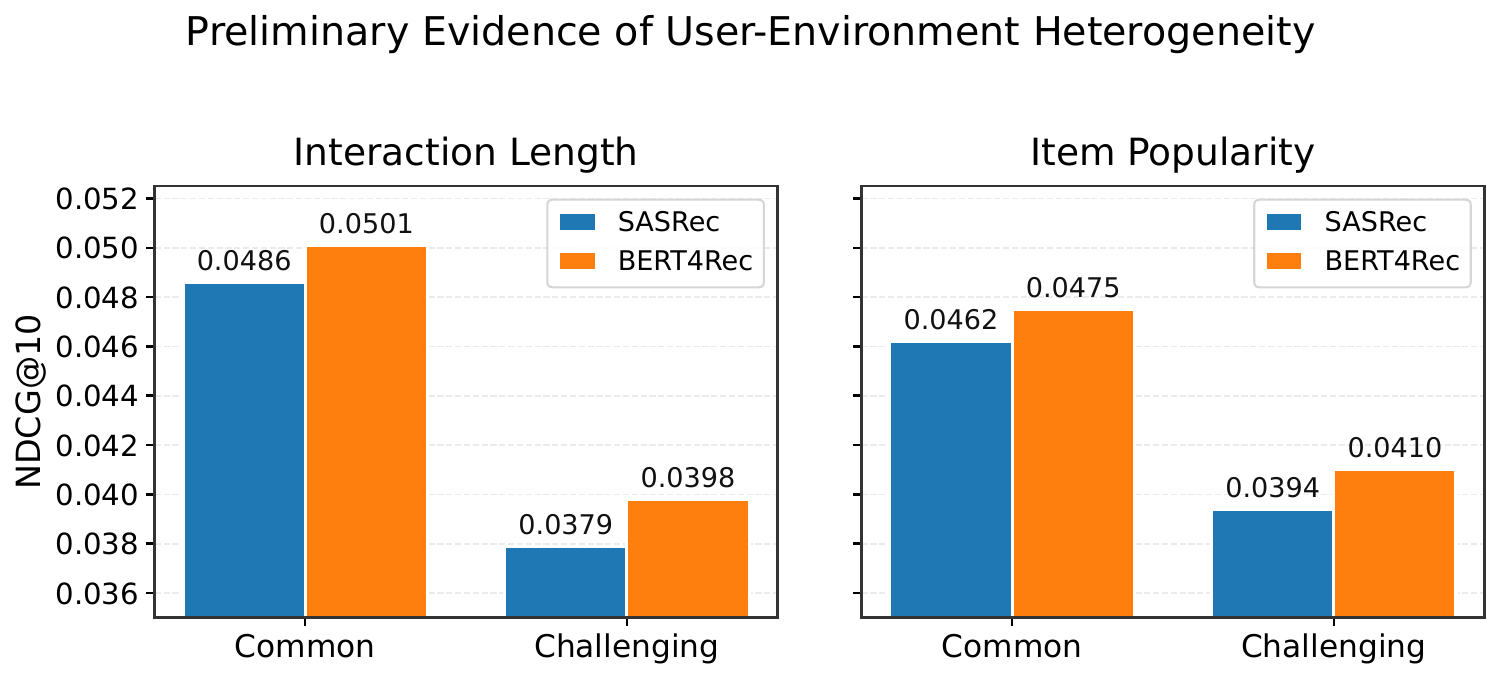}
    \caption{Preliminary evidence of user-environment heterogeneity under two partition views. Both \textsc{SASRec} and \textsc{BERT4Rec} exhibit substantially larger performance degradation in challenging environments, especially under the interaction-length partition.}
    \Description{The figure compares the performance of\textsc{SASRec} and \textsc{BERT4Rec} on common and challenging user groups under interaction-length and item-popularity partitions.}
    \label{fig:preliminary_heterogeneity}
\end{figure}

\section{Related Work}

\subsection{Sequential Recommender Systems}

Sequential recommender systems aim to model users' evolving interests. Early neural methods include recurrent and attention-based session models~\cite{hidasi2015session,quadrana2017personalizing,li2017neural,liu2018stamp}, while convolutional architectures such as \textsc{Caser}~\cite{tang2018personalized} and \textsc{NextItNet}~\cite{yuan2019simple} model local and high-order transitions. Transformer-based models, exemplified by \textsc{SASRec}~\cite{kang2018self}, \textsc{BERT4Rec}~\cite{sun2019bert4rec}, \textsc{LightSANs}~\cite{fan2021lighter}, and \textsc{FEARec}~\cite{du2023frequency}, are now strong next-item prediction backbones. Graph-based and all-MLP variants further broaden this model family~\cite{wu2019session,chang2021sequential,zhou2022filter}. Recent empirical work examines when modality-based representations are competitive with ID-based recommenders~\cite{yuan2023idmodality}, while other work combines next-item prediction and masked-language-modeling objectives to improve sequential recommendation~\cite{li2026metasr}.

Another important line improves representation learning through self-supervision, contrastive learning, and transfer learning. \textsc{S$^3$-Rec}~\cite{zhou2020s3rec}, \textsc{CL4SRec}~\cite{xie2022contrastive}, \textsc{DuoRec}~\cite{qiu2022contrastive}, \textsc{CBiT}~\cite{du2022contrastive}, and \textsc{ICLRec}~\cite{chen2022intent} improve sequence representations through auxiliary supervision. Adapter-based transfer learning has also been studied for parameter-efficient recommendation adaptation~\cite{fu2024adapter}. Language and multimodal methods further enrich item representations, improve robustness, and seek a better accuracy--efficiency balance~\cite{geng2022recommendation,hou2022unisrec,li2023text,li2024multi,xu2024slmrec,fu2024iisan,fu2025efficient,zhuang2025bridging,zhuang2025frequency,li2025teach}. Recent reproducibility work further shows that the contribution of multimodal embeddings can depend on the model architecture and fusion design~\cite{ye2026repro}. These studies primarily modify the representation learner or training objective. In contrast, \textsc{DS-Frame} leaves the backbone prediction pathway intact and studies whether an individual test sample should receive additional inference-time refinement.

\subsection{Adaptive Computation}

Conditional computation dynamically activates parts of a model according to the input, enabling an accuracy--efficiency trade-off~\cite{bengio2013estimating,bengio2017conditional,scardapane2024conditional}. Adaptive Computation Time (\textsc{ACT}) learns how many recurrent updates to perform~\cite{graves2016adaptive}; early-exit models stop at an intermediate depth when a prediction is sufficiently reliable~\cite{teerapittayanon2016branchynet,huang2018multi}; \textsc{SkipNet} learns to bypass selected layers~\cite{wang2018skipnet}; and mixture-of-experts (\textsc{MoE}) models conditionally activate a sparse set of experts~\cite{shazeer2017outrageously,fedus2022switch}.

\textsc{DS-Frame} shares the conditional-computation objective but operates at a different granularity. Rather than halting within a single computation graph, skipping layers, or selecting experts, it first reuses a backbone representation and then makes one sample-level decision between a Fast path and a fixed-step Slow refinement path. The selector is trained to estimate the marginal value of this additional refinement under a target routing budget; it therefore allocates a scarce slow-path budget across user sequences instead of only minimizing the computation of each sequence independently.

\subsection{Heterogeneous Users and Popularity Effects}

Recommendation quality can vary substantially with user behavior and the head--tail composition of historical interactions. Long-tail sequential recommendation studies explicitly seek to improve modeling for infrequent items or sparse users~\cite{hu2022memory,dang2026tail}, while behavioral analyses show that a single user model can fail to capture important differences across users~\cite{Ungruh2025monolith}. Recommendation algorithms may overexpose popular items and underexpose niche items~\cite{abdollahpouri2019managing}.

Our historical-item-popularity partition is not a popularity-debiasing intervention and \textsc{DS-Frame} does not claim to mitigate exposure or fairness bias. Instead, it uses the popularity profile of a user's observed history, together with interaction length, as an operational view of user environment and prediction difficulty. This distinction positions \textsc{DS-Frame} as adaptive computation for heterogeneous user environments, complementary to methods that directly optimize long-tail exposure or popularity fairness.

\subsection{Dual System Theory and Reasoning-Enhanced Recommendation}

Dual-System Theory (DST) distinguishes a fast, intuitive system from a slow, deliberative system~\cite{james1890principles,tversky1974judgment,kahneman2003maps}. It has motivated hybrid reasoning-and-retrieval frameworks in language processing~\cite{sun2025dual,cheng2025dualrag} and recent recommendation studies that distinguish immediate engagement from longer-term utility or model user decision processes~\cite{agarwal2024system,xu2023dual}.

Most closely related to our work are recent latent-reasoning~\cite{hao2025training,saunshi2025reasoning,geiping2026scaling,du2026latent,ye2026thinking} approaches for sequential recommendation. \textsc{ReaRec}~\cite{tang2025think}, \textsc{SlowRec}~\cite{zhang2025slow}, \textsc{LARES}~\cite{liu2025lares}, \textsc{PLR}~\cite{tang2026parallel}, and \textsc{ManCAR}~\cite{yang2026mancar} show that increasing test-time reasoning or refining hidden states can improve recommendation quality, especially for complex preference patterns. However, these methods primarily focus on improving the reasoning process itself (e.g., generating stronger reasoning trajectories or adaptively deciding when to stop within a single reasoning module), but cannot flexibly switch between different reasoning modes. \textsc{DS-Frame} is complementary: it explicitly separates a fast backbone path and a slow refinement path, then learns when slow computation is worthwhile for each sample.

\section{Preliminary Study: User-Environment Heterogeneity}
\label{sec:prelim_study}

Sequential recommendation models typically employ a static inference strategy, implicitly treating user environments as homogeneous. In this section, we conduct a preliminary empirical study to examine whether this assumption holds in practice. Specifically, we ask whether standard sequential recommenders behave consistently across common and operationally challenging user environments.

To answer this question, we segment users from two complementary behavioral views: interaction sequence length and historical item popularity. The former reflects the complexity of modeling long and potentially mixed-interest histories, while the latter reflects whether a user's historical preference concentrates on mainstream or long-tail items. We then evaluate the performance on the resulting user groups using two representative sequential recommenders, \textsc{SASRec}~\cite{kang2018self} and \textsc{BERT4Rec}~\cite{sun2019bert4rec}.

\subsection{Empirical Study}
\label{sec:preliminary_heterogeneity}

For each dataset, we train each backbone once on the full training set and evaluate it separately on different user groups in the test set. This protocol keeps the training data and model parameters unchanged, so any observed gap directly reflects how the same static recommender behaves under different user environments.

We construct user groups from two complementary views.

\paragraph{Interaction Length.}
We use interaction length as an operational proxy for prediction difficulty. For each dataset $d$, we determine a dataset-specific threshold $L_d^*$ from the training-user length distribution. Specifically, we construct the complementary cumulative distribution function (CCDF) of interaction lengths and identify an elbow point on the log-CCDF curve via piecewise linear fitting. To improve robustness against extremely rare tail lengths, the elbow search is restricted to the central mass of the distribution. Users with $L_u \leq L_d^*$ are assigned to the common environment, whereas users with $L_u > L_d^*$ are assigned to the challenging environment.

\paragraph{Item Popularity.}
We further use the average item popularity in a user's history as a proxy for taste mainstreamness. Users with higher average popularity are assigned to the common environment, whereas those with lower average popularity are assigned to the challenging environment. For a clearer contrast, we set $p=20$ for all datasets and compare the top-20\% and bottom-20\% user groups under this ranking.

\begin{table}[t]
\centering
\caption{Preliminary evidence of user-environment heterogeneity. We report NDCG@10 under two partition views; Drop denotes the relative decrease from common to challenging environments. Interaction-length ratios are aggregated across all datasets.}
\label{tab:preliminary_evidence}
\setlength{\tabcolsep}{4.5pt}
\renewcommand{\arraystretch}{1.12}
\resizebox{\columnwidth}{!}{
\begin{tabular}{llcccc}
\toprule
View & Backbone & Common & Challenging & Drop & User Ratio \\
\midrule
\multirow{2}{*}{Interaction Length}
& \textsc{SASRec}   & 0.0486 & 0.0379 & 22.0\% & 86.2:13.8 \\
& \textsc{BERT4Rec} & 0.0501 & 0.0398 & 20.6\% & 86.2:13.8 \\
\midrule
\multirow{2}{*}{Item Popularity}
& \textsc{SASRec}   & 0.0462 & 0.0394 & 14.7\% & 20:20 \\
& \textsc{BERT4Rec} & 0.0475 & 0.0410 & 13.7\% & 20:20 \\
\bottomrule
\end{tabular}
}
\end{table}

Based on the results in Figure~\ref{fig:preliminary_heterogeneity} and Table~\ref{tab:preliminary_evidence}, we have the following observations:
\begin{itemize}[leftmargin=*, topsep=0pt]
    \item Both \textsc{SASRec} and \textsc{BERT4Rec} consistently perform better in the common environment than in the challenging environment under both partition views.
    \item The gap is particularly pronounced under the interaction-length view, where the performance drops over 20\% for both backbones.
\end{itemize}
These observations indicate that static one-pass recommenders behave unevenly across heterogeneous user environments. We note that the challenging group is smaller under the interaction-length partition, because long-history users naturally lie in the tail of the user-length distribution. Therefore, this preliminary study is not intended as a standalone statistical claim. Instead, it provides motivational evidence of user-environment heterogeneity: the degradation is consistent for both \textsc{SASRec} and \textsc{BERT4Rec}, and a similar trend also appears under the item-popularity partition, where the compared groups are balanced by construction. This motivates \textsc{DS-Frame}, an adaptive fast--slow framework that allocates additional computation when it is likely to improve prediction. We further validate this trend through the group-wise gain and routing analyses in Sections \ref{sec:groupwise_gain} and \ref{sec:selector_analysis}.

This gap suggests that static recommenders may allocate computation suboptimally across heterogeneous users. It motivates \textsc{DS-Frame}, an adaptive fast and slow framework that allocates additional computation when it is likely to improve prediction.

\section{Problem Formulation}
\label{sec:preliminary}

\subsection{Sequential Recommendation}
\label{subsec:problem_definition}

Let $U$ denote the set of $M$ users and $V$ denote the set of $N$ items. Each item $v \in V$ is associated with an embedding vector $\mathbf{e}_v \in \mathbb{R}^d$, where $d$ is the embedding dimension. For each user $u \in U$, the interaction history is represented as a chronological sequence
\begin{equation}
S_u = (v_{u,1}, v_{u,2}, \dots, v_{u,n_u}),
\end{equation}
where $v_{u,t} \in V$ is the item interacted with at step $t$, and $n_u = |S_u|$ is the sequence length.

Given the prefix sequence
\begin{equation}
S_{u,1:t} = (v_{u,1}, v_{u,2}, \dots, v_{u,t}),
\end{equation}
the goal of sequential recommendation is to predict the next item $v_{u,t+1}$ that the user will interact with.

Let $\Theta$ denote the parameters of a recommendation model. The model takes the historical sequence as input and estimates the probability of the next item:
\begin{equation}
P(v_{u,t+1}\mid S_{u,1:t};\Theta).
\end{equation}
The standard training objective is to maximize the log-likelihood of the ground-truth next item over all users and valid sequence positions:
\begin{equation}
\max_{\Theta}
\sum_{u \in U}\sum_{t=1}^{n_u-1}
\log P(v_{u,t+1}\mid S_{u,1:t};\Theta).
\label{eq:objective_prob}
\end{equation}

\begin{figure}[t]
    \centering
    \includegraphics[width=\linewidth]{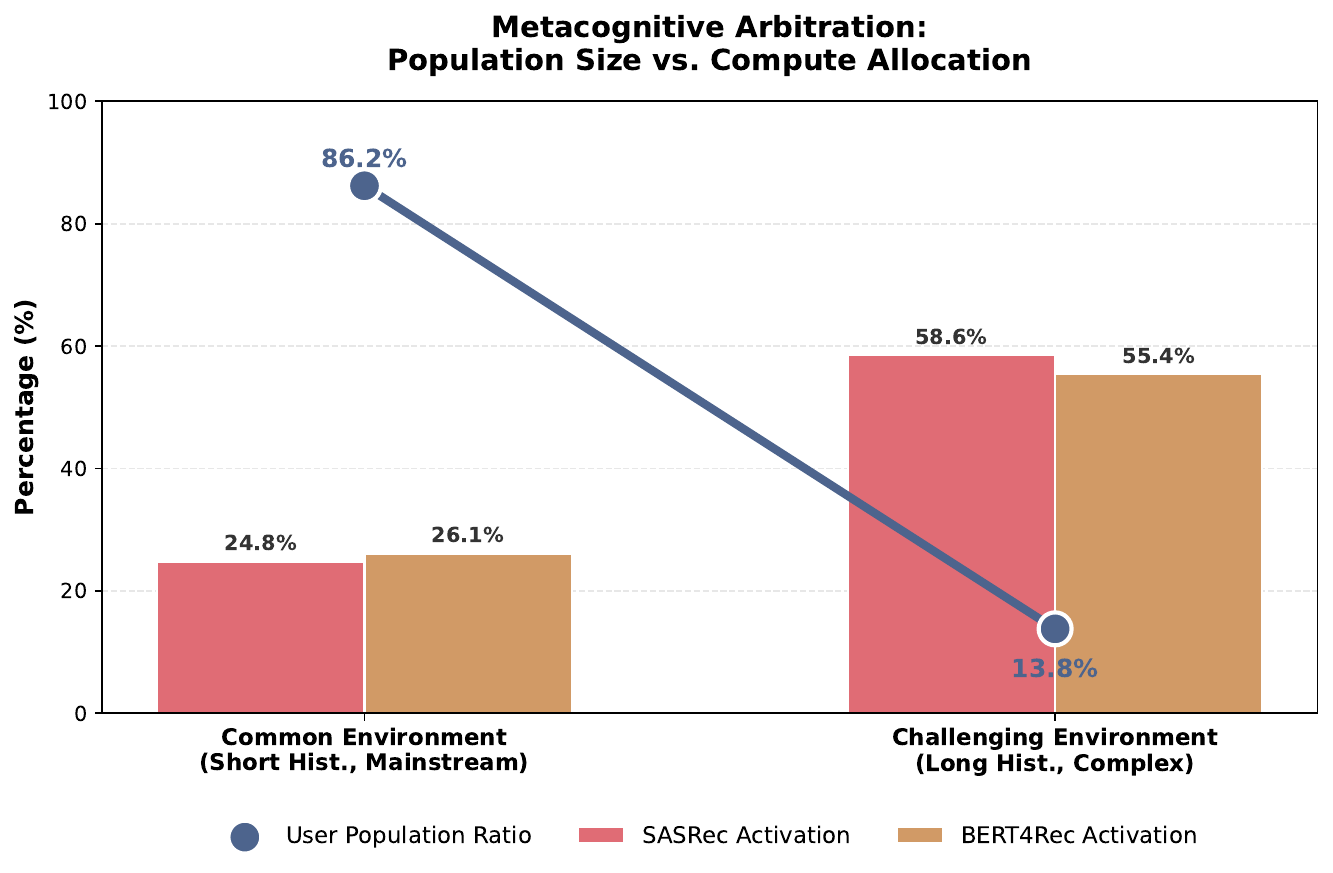}
    \caption{Slow-System activation behavior.}
    \Description{The figure visualizes the activation behavior of the Slow System under the learned selector.}
    \label{fig:activation}
\end{figure}

\subsection{Accuracy--Efficiency Trade-off}
\label{sec:preliminary_tradeoff}

Sequential recommendation models must balance two competing goals: \emph{predictive accuracy} and \emph{computational efficiency}. Stronger models often benefit from richer contextual information, improving prediction quality especially for users with long, noisy, or mixed-interest histories. However, preserving and processing more information requires heavier computation, which is undesirable in large-scale, latency-sensitive recommendation systems.

This trade-off can be viewed through the lens of the Information Bottleneck (IB) principle. Conceptually, a recommendation model learns a latent representation $Z$ from the input sequence $X$ to predict the target item $Y$. A more compressed representation is computationally efficient but may omit useful details, whereas a less compressed one may preserve more predictive information at higher cost. Thus, sequential recommendation naturally involves a tension between compression and predictive sufficiency.

In this work, we do not directly optimize an IB objective. Instead, we use this perspective to motivate our framework design. The \emph{Fast System} serves as a lightweight pathway with compact computation, while the \emph{Slow System} performs additional iterative reasoning to preserve task-relevant information for challenging cases. This motivates adaptive routing that decides when the extra cost of slow reasoning is worthwhile.
\begin{figure*}[t]
    \centering
    \includegraphics[
        width=\textwidth,
        trim=2 170 2 90,
        clip
    ]{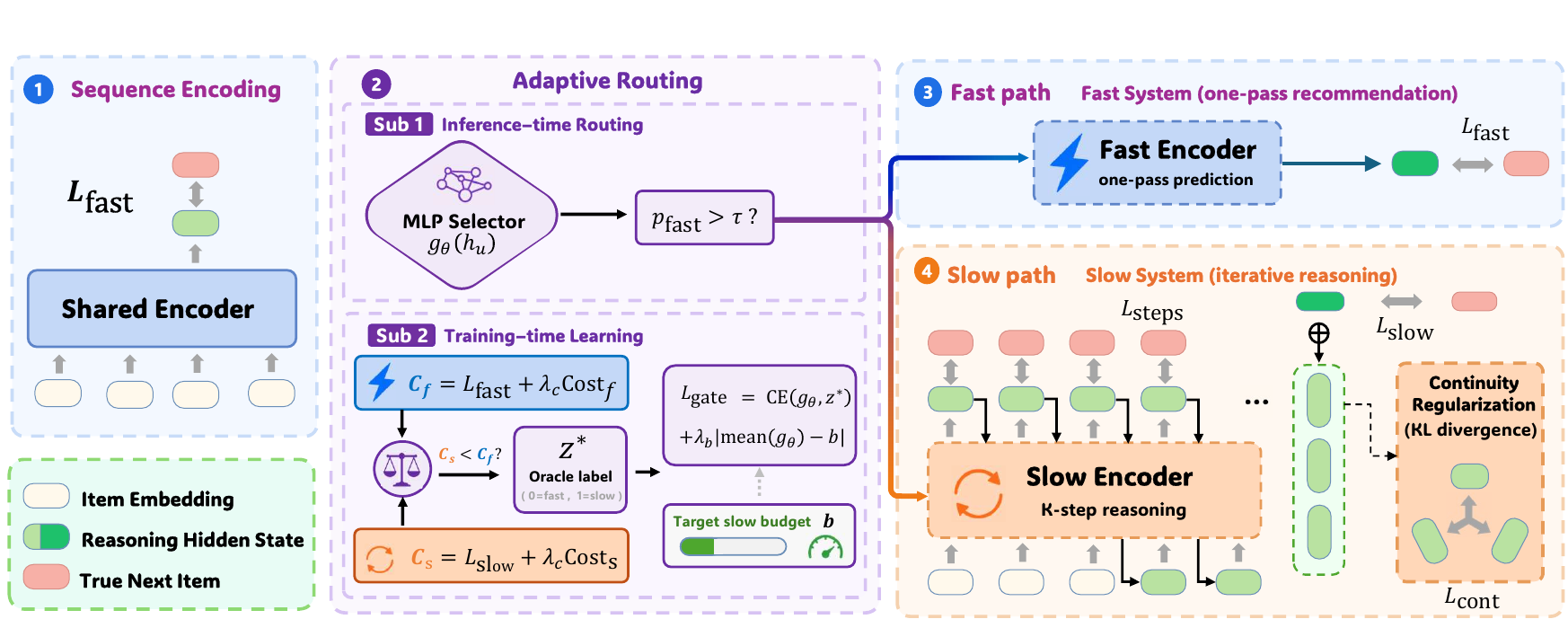}
    \caption{Overall architecture of \textsc{DS-Frame}. A shared encoder extracts the user representation, and a selector routes each sample to the Fast or Slow System. The selector is trained with oracle-guided, budget-regularized supervision, while the Slow System performs $K$-step latent refinement with multi-step and continuity regularization.}
    \Description{The figure shows the overall\textsc{DS-Frame} architecture, including a shared sequence encoder, a Fast System, a Slow System, and a selector gate that routes user sequences to different computational pathways.}
    \label{fig:main}
\end{figure*}

\section{Methodology}
\label{sec:methodology}

\textsc{DS-Frame} is an adaptive fast--slow inference framework designed to balance efficiency and accuracy in sequential recommendation. Given a user's historical interaction sequence $S_u = (v_{u,1}, v_{u,2}, \ldots, v_{u,n_u})$, \textsc{DS-Frame} first builds a shared sequence representation, which is then used by two pathways: a \emph{Fast System} for one-pass prediction and a \emph{Slow System} for iterative latent refinement. A lightweight \emph{Selector Gate} determines whether an input should use the Fast System or invoke the Slow System.

Formally, each item $v_{u,i}$ in $S_u$ is first mapped to an input representation:
\begin{equation}
h_i^0 = e_{v_{u,i}} + p_i,
\label{eq:input_representation}
\end{equation}
where $e_{v_{u,i}} \in \mathbb{R}^d$ is the item embedding and $p_i \in \mathbb{R}^d$ is a learnable positional embedding. This yields an initial sequence representation
\begin{equation}
H^0 = (h_1^0, h_2^0, \ldots, h_{n_u}^0).
\label{eq:initial_sequence}
\end{equation}

An $L$-layer Transformer encoder, denoted by $f_{\text{Enc}}$, processes $H^0$ as
\begin{equation}
H^L = f_{\text{Enc}}(H^0),
\label{eq:transformer_output}
\end{equation}
where $H^L = (h_1^L, h_2^L, \ldots, h_{n_u}^L)$ is the final hidden sequence. The last hidden state
\begin{equation}
h_u = H^L[n_u]
\label{eq:shared_user_representation}
\end{equation}
serves as the shared user representation for both systems.

Built on top of this shared encoder, \textsc{DS-Frame} combines a Fast System for efficiency, a Slow System for precision, and a selector gate for adaptive routing. During training, both the Fast and Slow Systems are evaluated to construct a sample-level routing target for the selector. During inference, the learned selector alone determines whether an input is processed by the Fast or Slow pathway. The overall architecture is illustrated in Figure~\ref{fig:main}.

\subsection{Fast Thinking System (System 1)}
\label{subsec:fast_thinking_system}

The Fast System is designed for routine recommendation scenarios where a compact user representation is sufficient. It directly uses the shared representation $h_u$ from Eq.~\ref{eq:shared_user_representation} to predict the next item:
\begin{equation}
\hat{\mathbf{y}}^{f} = W_f h_u + b_f,
\label{eq:fast_logits}
\end{equation}
where $\hat{\mathbf{y}}^{f} \in \mathbb{R}^{|V|}$ denotes the prediction logits over the item set and $W_f, b_f$ are trainable parameters.

The Fast System is optimized with the standard next-item prediction loss:
\begin{equation}
\mathcal{L}_{\mathrm{fast}} = \ell(\hat{\mathbf{y}}^{f}, y),
\label{eq:l_fast}
\end{equation}
where $y$ is the ground-truth next item and $\ell(\cdot,\cdot)$ denotes the recommendation loss, e.g., cross-entropy over the target item.

Conceptually, the Fast System provides a computationally efficient pathway for samples where one-pass inference is already sufficient.

\subsection{Slow Thinking System (System 2)}
\label{subsec:slow_thinking_system}

The Slow System is designed for samples where one-pass inference may be insufficient, such as users with longer histories or less mainstream item profiles. It operates on the full shared sequence representation $H^L$ and performs iterative latent refinement over $K$ steps. Let
\begin{equation}
H_{\mathrm{seq}} = H^L, \qquad r^{(0)} = H_{\mathrm{seq}}[n_u].
\label{eq:slow_init}
\end{equation}
At each reasoning step $k \in \{1,\dots,K\}$, a reasoning token is formed as
\begin{equation}
x_{\mathrm{reason}}^{(k)} = r^{(k-1)} + p_k^R,
\label{eq:thinking_unit}
\end{equation}
where $p_k^R \in \mathbb{R}^d$ is a learnable step embedding.

We then augment the original sequence with all reasoning tokens generated so far:
\begin{equation}
H_{\mathrm{aug}}^{(k)} =
\left[ H_{\mathrm{seq}};\, x_{\mathrm{reason}}^{(1)};\, \cdots ;\, x_{\mathrm{reason}}^{(k)} \right],
\label{eq:augmented_sequence}
\end{equation}
where $[\cdot;\cdot]$ denotes sequence concatenation. This design allows each reasoning step to access both the original interaction history and the evolving intermediate reasoning states.

The augmented sequence is processed by a reasoning block:
\begin{equation}
\widetilde{H}^{(k)} = f_{\mathrm{reason}}(H_{\mathrm{aug}}^{(k)}),
\label{eq:reason_block}
\end{equation}
where $f_{\mathrm{reason}}$ is a Transformer-style reasoning module whose parameters are shared across reasoning steps for efficiency. The refined representation at step $k$ is taken as the output corresponding to the last reasoning token:
\begin{equation}
r^{(k)} = \widetilde{H}^{(k)}[-1].
\label{eq:refined_representation_k}
\end{equation}

Through $K$ iterative steps, the Slow System produces a sequence of refined representations
\[
\{r^{(0)}, r^{(1)}, \ldots, r^{(K)}\}.
\]

To stabilize and guide this multi-step reasoning process, we introduce three training components.

\paragraph{Multi-step Reasoning Supervision.}
Each intermediate representation $r^{(k)}$ is supervised to predict the target item:
\begin{equation}
\hat{\mathbf{y}}^{(k)} = W_s r^{(k)} + b_s,
\label{eq:slow_step_logits}
\end{equation}
\begin{equation}
\mathcal{L}_{\mathrm{steps}} =
\sum_{k=0}^{K}
\ell\!\left(
\frac{\hat{\mathbf{y}}^{(k)}}{\tau_k}, y
\right),
\label{eq:multi_step_loss}
\end{equation}
where $\tau_k$ is the temperature used at step $k$.

\paragraph{Continuity Regularization.}
To encourage smooth transitions between adjacent reasoning steps, we regularize the output distributions of neighboring steps:
\begin{equation}
\mathcal{L}_{\mathrm{cont}} =
\sum_{k=0}^{K-1}
\mathrm{KL}
\left(
P(\cdot \mid r^{(k)}; \tau_k)
\;\|\;
P(\cdot \mid r^{(k+1)}; \tau_{k+1})
\right).
\label{eq:continuity_loss}
\end{equation}

\paragraph{Progressive Temperature Annealing.}
We use a step-dependent temperature schedule:
\begin{equation}
\tau_k = \tau_{\mathrm{base}} \cdot \alpha^{K-k},
\label{eq:pta_temperature}
\end{equation}
where $\tau_{\mathrm{base}} > 0$ and $\alpha > 0$ are hyperparameters. This schedule allows the early reasoning steps to remain softer and the later steps to become more decisive.

For final prediction, we aggregate the logits from all reasoning steps by averaging:
\begin{equation}
\hat{\mathbf{y}}^{s}
=
\frac{1}{K+1}\sum_{k=0}^{K}\hat{\mathbf{y}}^{(k)}.
\label{eq:slow_final_logits}
\end{equation}
Accordingly, the Slow System is optimized by
\begin{equation}
\mathcal{L}_{\mathrm{slow}}
=
\ell(\hat{\mathbf{y}}^{s}, y)
+
\lambda_{\mathrm{step}} \mathcal{L}_{\mathrm{steps}}
+
\lambda_{\mathrm{cont}} \mathcal{L}_{\mathrm{cont}},
\label{eq:l_slow}
\end{equation}
where $\lambda_{\mathrm{step}}$ and $\lambda_{\mathrm{cont}}$ control the strengths of the auxiliary reasoning losses.

Compared with the Fast System, the Slow System allocates additional computation to perform iterative refinement over the sequence representation. It is suitable for challenging cases where a single-pass representation may be insufficient.

\subsection{Selector Training Objective: Cost-Aware Routing}
\label{sec:selector_objective}

The key challenge in \textsc{DS-Frame} is to learn a selector that decides, for each input sequence, whether the Fast System is sufficient or whether the Slow System should be invoked. We formulate this routing problem as a cost-aware binary decision problem.

Let $S_f$ and $S_s$ denote the Fast and Slow Systems, respectively. For an input-target pair $(x,y)$, we define the sample-level total cost of invoking system $S \in \{S_f, S_s\}$ as
\begin{equation}
\mathcal{C}(S; x,y) = \ell(S(x), y) + \lambda_{\mathrm{c}} \cdot \mathrm{Cost}(S),
\label{eq:system_cost}
\end{equation}
where $\ell(S(x), y)$ is the prediction loss of system $S$ on sample $(x,y)$, $\mathrm{Cost}(S)$ denotes the normalized inference cost of invoking system $S$, and $\lambda_{\mathrm{c}} \ge 0$ controls the accuracy--efficiency trade-off. In our implementation, we use a step-based normalized cost proxy: the Fast System is assigned $\mathrm{Cost}(S_f)=1$, while the Slow System reuses the shared encoder output and performs $K$ additional latent-refinement steps, so $\mathrm{Cost}(S_s)=1+\gamma K$. Unless otherwise specified, we set $\gamma=1$ as an architecture-agnostic proxy for one refinement step. This normalized cost is used for constructing training-time oracle routing labels and budget-aware routing analysis.

Based on Eq.~\ref{eq:system_cost}, the ideal routing decision is to select the system with lower total cost. This yields the oracle routing label
\begin{equation}
z^*(x,y) =
\mathbb{I}
\big(
\mathcal{C}(S_s; x,y) < \mathcal{C}(S_f; x,y)
\big),
\label{eq:oracle_label}
\end{equation}
where $z^*(x,y)=1$ indicates that the sample should be routed to the Slow System, and $z^*(x,y)=0$ indicates that the Fast System is sufficient.

The selector is implemented as a lightweight MLP gate:
\begin{equation}
g_\theta(x) = \sigma\!\left( W_2\, \phi(W_1 h_u + b_1) + b_2 \right),
\label{eq:gate_function}
\end{equation}
where $g_\theta(x)\in(0,1)$ is the predicted probability of routing $x$ to the Slow System, $\phi(\cdot)$ is a non-linear activation function, and $\sigma(\cdot)$ is the sigmoid function.

Since the oracle decision in Eq.~\ref{eq:oracle_label} depends on sample-level losses and is therefore only available during training, we use it as a pseudo-supervision signal for learning the selector. At inference time, the selector does not access ground-truth labels or sample-level losses; it only uses the learned mapping from the input representation to the routing decision.

\subsection{Practical Gate Loss with Budget Regularization}
\label{sec:practical_gate_loss}

Based on the oracle routing label, we train the selector using the following guidance loss:

\begin{equation}
\begin{aligned}
\mathcal{L}_{\mathrm{guide}}
= -\frac{1}{|\mathcal{B}|}
\sum_{(x,y)\in\mathcal{B}}
\Big[
& z^*(x,y)\log g_\theta(x) \\
& + \big(1-z^*(x,y)\big)
\log\big(1-g_\theta(x)\big)
\Big],
\end{aligned}
\label{eq:l_guide}
\end{equation}

where $\mathcal{B}$ is the current mini-batch. This term encourages the selector to imitate the oracle routing rule.

To explicitly control how often the Slow System is activated, we further introduce a budget regularizer:
\begin{equation}
\mathcal{L}_{\mathrm{budget}}
=
\left|
\frac{1}{|\mathcal{B}|}\sum_{x\in\mathcal{B}} g_\theta(x) - b
\right|,
\label{eq:l_budget}
\end{equation}
where $b\in(0,1)$ is a target slow-routing ratio. This term constrains the selector's average activation rate to stay close to a desired computational budget.

The final selector loss is defined as
\begin{equation}
\mathcal{L}_{\mathrm{gate}}
=
\mathcal{L}_{\mathrm{guide}}
+
\lambda_{\mathrm{b}}\mathcal{L}_{\mathrm{budget}},
\label{eq:l_gate}
\end{equation}
where $\lambda_{\mathrm{b}}$ controls the strength of budget regularization.

Together, $\mathcal{L}_{\mathrm{guide}}$ and $\mathcal{L}_{\mathrm{budget}}$ guide \textsc{DS-Frame} to allocate slow refinement selectively, rather than applying the Slow System uniformly to all samples.

\subsection{Overall Training and Inference}
\label{subsec:overall_training}

The overall training objective of \textsc{DS-Frame} combines the Fast-System loss, the Slow-System loss, and the selector loss:
\begin{equation}
\mathcal{L}_{\mathrm{total}}
=
\mathcal{L}_{\mathrm{fast}}
+
\mathcal{L}_{\mathrm{slow}}
+
\lambda_{\mathrm{g}}\mathcal{L}_{\mathrm{gate}},
\label{eq:l_total}
\end{equation}
where $\lambda_{\mathrm{g}}$ balances recommendation learning and routing learning.

During training, both systems are evaluated on each sample to obtain their sample-level losses, which are then used to construct the oracle routing label in Eq.~\ref{eq:oracle_label}. The selector is trained to imitate this oracle while respecting the global routing budget.

During inference, only the learned selector is used. Specifically, given an input sequence $x$, we compute the gate score $g_\theta(x)$ and route the sample according to
\begin{equation}
\text{Route}(x)=
\begin{cases}
S_s, & \text{if } g_\theta(x) > \tau,\\
S_f, & \text{otherwise},
\end{cases}
\label{eq:inference_routing}
\end{equation}
where $\tau \in (0,1)$ is a routing threshold. In this way, \textsc{DS-Frame} achieves an adaptive trade-off between efficiency and precision.

\section{Experiments}
\label{sec:experiments}

In this section, we conduct experiments to answer the following Research Questions (\textbf{RQ}s):
\begin{itemize}[leftmargin=*, topsep=0pt]
    \item \textbf{RQ1}: Can \textsc{DS-Frame} achieve competitive overall recommendation performance?
    \item \textbf{RQ2}: Does \textsc{DS-Frame} achieve larger gains on challenging user environments, where static backbones suffer more severe performance degradation?
    \item \textbf{RQ3}: Does the learned selector perform effective routing by allocating slow-path computation more appropriately than random or uniform routing strategies?
    \item \textbf{RQ4}: How does \textsc{DS-Frame} trade off recommendation accuracy and computational cost under different routing budgets?
    \item \textbf{RQ5}: How does each key component of \textsc{DS-Frame} contribute to the final performance?
\end{itemize}

\subsection{Experimental Setup}
\label{sec:experimental_setup}

\subsubsection{Datasets}

We evaluate \textsc{DS-Frame} on five real-world sequential recommendation datasets from two sources: one Yelp dataset and four Amazon domains, namely \textit{Video Games}, \textit{Beauty}, \textit{Sports}, and \textit{Toys}. We use chronological user--item interactions, apply standard filtering for data quality, and split interactions by timestamp to avoid temporal leakage.

\begin{enumerate}[leftmargin=*, topsep=0pt]
    \item \textbf{Yelp:} A business review dataset containing chronological user--business interactions\footnote{Yelp Open Dataset: \url{https://business.yelp.com/data/resources/open-dataset/}}.
    \item \textbf{Amazon 2023:} Four domain-specific datasets derived from \textit{Amazon Review}~\cite{hou2024bridging}, including \textit{Video Games}, \textit{Beauty}, \textit{Sports \& Outdoors}, and \textit{Toys \& Games}. We use their chronological user--item interactions for sequential recommendation.
\end{enumerate}

\subsubsection{Evaluation Metrics}

We report standard top-$k$ ranking metrics, including \textbf{Normalized Discounted Cumulative Gain at 10 (NDCG@10)}, \textbf{NDCG@20}, \textbf{Hit Rate at 10 (HR@10)}, and \textbf{HR@20} ~\cite{chang2021sequential}. NDCG evaluates both the correctness and ranking quality of the recommended items, while HR measures whether the ground-truth item appears in the top-$k$ recommendation list. We adopt full-ranking evaluation: for each test instance, the ground-truth next item is ranked against all candidate items not present in the user's training history, instead of sampled negative items.

\subsubsection{Backbone Models}

We instantiate \textsc{DS-Frame} on two representative sequential recommendation backbones and compare each \textsc{DS-Frame} variant with its original backbone version:
\begin{enumerate}[leftmargin=*, topsep=0pt]
    \item \textbf{SASRec:} A causal sequential recommendation model based on a unidirectional Transformer~\cite{kang2018self}.
    \item \textbf{BERT4Rec:} A bidirectional sequential recommendation model based on the Transformer encoder~\cite{sun2019bert4rec}. For next-item evaluation with \textsc{BERT4Rec}, we use the hidden state at the appended prediction position to score candidate items.
\end{enumerate}

\subsubsection{Baselines}

To comprehensively evaluate \textsc{DS-Frame}, we compare it with three groups of baselines. \textbf{Conventional sequential recommenders} include \textsc{SASRec}~\cite{kang2018self} and \textsc{BERT4Rec}~\cite{sun2019bert4rec}, which follow the standard one-pass inference paradigm. \textbf{Reasoning-enhanced sequential recommenders} include \textsc{ReaRec-ERL} and \textsc{ReaRec-PRL}~\cite{tang2025think}, which extend sequential recommenders with implicit multi-step latent reasoning, as well as \textsc{STREAM-Rec}~\cite{zhang2025slow}, which adopts an explicit slow-thinking paradigm via iterative reasoning generation. We further include stronger recent latent-reasoning methods, namely \textsc{LARES}~\cite{liu2025lares} and \textsc{ManCAR}~\cite{yang2026mancar}, which improve reasoning quality through latent representation refinement and adaptive test-time computation.

\subsubsection{Implementation Details}

We conduct all experiments on a single NVIDIA A100 GPU. Unless otherwise specified, we follow the standard implementations of the backbone models and keep their architectures unchanged. We set the embedding size to 256, batch size to 128, learning rate to 0.001, use Adam as the optimizer, GeLU as the activation function, and truncate user sequences to a maximum length of 50. Early stopping is applied if the validation metric does not improve for 10 consecutive epochs.

For \textsc{DS-Frame}, the Slow System is trained jointly with the backbone objective, multi-step supervision loss, continuity regularization term, and selector loss. We tune the number of reasoning steps $K$ over $\{1,2,3\}$, the base temperature $\tau_{\mathrm{base}}$ over $\{0.05, 0.1, 0.5, 1.0, 2.0\}$, and the decay rate $\alpha$ over $\{1.0, 1.2, 1.5, 2.0, 5.0\}$. The loss weights $\lambda_{\mathrm{step}}$, $\lambda_{\mathrm{cont}}$, $\lambda_g$, and $\lambda_b$ are tuned over $\{0.001, 0.005, 0.01, 0.05, 0.1\}$. We tune the routing budget $b$ over $\{0.2, 0.4, 0.6\}$ and the routing threshold $\tau_{\mathrm{route}}$ over $\{0.3, 0.4, 0.5, 0.6, 0.7\}$ on the validation set.

For fair comparison, each \textsc{DS-Frame} variant uses the same backbone as its corresponding baseline, including \textsc{SASRec} and \textsc{BERT4Rec}. We select all hyperparameters according to validation performance and report the best test results.

\subsection{Main Performance Comparison (RQ1)}
\label{sec:main_results}
Table~\ref{tab:main_results} reports the main recommendation performance of \textsc{DS-Frame} on two representative sequential recommendation backbones, \textsc{SASRec} and \textsc{BERT4Rec}. Overall, \textsc{DS-Frame} consistently improves both backbones across all datasets and metrics, demonstrating the effectiveness and portability of the proposed dual-system framework.

For \textsc{SASRec}, \textsc{DS-Frame} achieves average relative improvements of \textbf{7.5\%}, \textbf{6.6\%}, \textbf{6.6\%}, and \textbf{6.2\%} on NDCG@10, NDCG@20, HR@10, and HR@20, respectively. For \textsc{BERT4Rec}, it yields average relative gains of \textbf{6.7\%}, \textbf{6.3\%}, \textbf{6.0\%}, and \textbf{5.9\%}. These improvements are consistent across all datasets in Table~\ref{tab:main_results}, rather than being concentrated on a single benchmark.

We further observe notable gains on \textbf{Beauty}, \textbf{Sports}, and \textbf{Toys}, where NDCG@10 improves by more than \textbf{7\%} on at least one backbone. This suggests that \textsc{DS-Frame} is especially effective when recommendation requires finer-grained modeling beyond a single static backbone.

These results answer \textbf{RQ1} positively: rather than replacing existing recommenders with a standalone architecture, \textsc{DS-Frame} serves as a plug-and-play enhancement framework that consistently strengthens strong sequential recommendation backbones.

\begin{table*}[t]
\centering
\caption{Main results of \textsc{DS-Frame} on two representative sequential recommendation backbones. $^\dagger$ marks improvements over the corresponding backbone.}
\label{tab:main_results}
\setlength{\tabcolsep}{3.2pt}
\renewcommand{\arraystretch}{1.0}
\resizebox{\textwidth}{!}{
\begin{tabular}{l|cccc|cccc|cccc|cccc}
\toprule
\multirow{3}{*}{Dataset}
& \multicolumn{8}{c|}{\textsc{SASRec}}
& \multicolumn{8}{c}{\textsc{BERT4Rec}} \\
\cmidrule(lr){2-9}\cmidrule(lr){10-17}
& \multicolumn{4}{c|}{Base}
& \multicolumn{4}{c|}{+ \textsc{DS-Frame}$^\dagger$}
& \multicolumn{4}{c|}{Base}
& \multicolumn{4}{c}{+ \textsc{DS-Frame}$^\dagger$} \\
\cmidrule(lr){2-5}\cmidrule(lr){6-9}\cmidrule(lr){10-13}\cmidrule(lr){14-17}
& N@10 & N@20 & HR@10 & HR@20
& N@10 & N@20 & HR@10 & HR@20
& N@10 & N@20 & HR@10 & HR@20
& N@10 & N@20 & HR@10 & HR@20 \\
\midrule

Video Games
& 0.0552 & 0.0796 & 0.1113 & 0.1587
& \makecell{\textbf{0.0590}\\[-1pt]{\scriptsize($\uparrow$6.9\%)}} 
& \makecell{\textbf{0.0845}\\[-1pt]{\scriptsize($\uparrow$6.2\%)}} 
& \makecell{\textbf{0.1178}\\[-1pt]{\scriptsize($\uparrow$5.8\%)}} 
& \makecell{\textbf{0.1669}\\[-1pt]{\scriptsize($\uparrow$5.2\%)}}
& 0.0569 & 0.0812 & 0.1130 & 0.1611
& \makecell{\textbf{0.0604}\\[-1pt]{\scriptsize($\uparrow$6.1\%)}} 
& \makecell{\textbf{0.0858}\\[-1pt]{\scriptsize($\uparrow$5.7\%)}} 
& \makecell{\textbf{0.1191}\\[-1pt]{\scriptsize($\uparrow$5.4\%)}} 
& \makecell{\textbf{0.1690}\\[-1pt]{\scriptsize($\uparrow$4.9\%)}} \\
  
Beauty
& 0.0418 & 0.0498 & 0.0813 & 0.1164
& \makecell{\textbf{0.0453}\\[-1pt]{\scriptsize($\uparrow$8.4\%)}} 
& \makecell{\textbf{0.0534}\\[-1pt]{\scriptsize($\uparrow$7.2\%)}} 
& \makecell{\textbf{0.0869}\\[-1pt]{\scriptsize($\uparrow$6.9\%)}} 
& \makecell{\textbf{0.1234}\\[-1pt]{\scriptsize($\uparrow$6.0\%)}}
& 0.0430 & 0.0512 & 0.0830 & 0.1192
& \makecell{\textbf{0.0461}\\[-1pt]{\scriptsize($\uparrow$7.2\%)}} 
& \makecell{\textbf{0.0550}\\[-1pt]{\scriptsize($\uparrow$7.4\%)}} 
& \makecell{\textbf{0.0884}\\[-1pt]{\scriptsize($\uparrow$6.5\%)}} 
& \makecell{\textbf{0.1266}\\[-1pt]{\scriptsize($\uparrow$6.2\%)}} \\

Sports
& 0.0226 & 0.0281 & 0.0480 & 0.0702
& \makecell{\textbf{0.0244}\\[-1pt]{\scriptsize($\uparrow$8.0\%)}} 
& \makecell{\textbf{0.0302}\\[-1pt]{\scriptsize($\uparrow$7.5\%)}} 
& \makecell{\textbf{0.0516}\\[-1pt]{\scriptsize($\uparrow$7.5\%)}} 
& \makecell{\textbf{0.0750}\\[-1pt]{\scriptsize($\uparrow$6.8\%)}}
& 0.0231 & 0.0286 & 0.0489 & 0.0715
& \makecell{\textbf{0.0248}\\[-1pt]{\scriptsize($\uparrow$7.4\%)}} 
& \makecell{\textbf{0.0306}\\[-1pt]{\scriptsize($\uparrow$7.0\%)}} 
& \makecell{\textbf{0.0522}\\[-1pt]{\scriptsize($\uparrow$6.7\%)}} 
& \makecell{\textbf{0.0758}\\[-1pt]{\scriptsize($\uparrow$6.0\%)}} \\

Toys
& 0.0435 & 0.0521 & 0.0668 & 0.1045
& \makecell{\textbf{0.0469}\\[-1pt]{\scriptsize($\uparrow$7.8\%)}} 
& \makecell{\textbf{0.0560}\\[-1pt]{\scriptsize($\uparrow$7.5\%)}} 
& \makecell{\textbf{0.0712}\\[-1pt]{\scriptsize($\uparrow$6.6\%)}} 
& \makecell{\textbf{0.1119}\\[-1pt]{\scriptsize($\uparrow$7.1\%)}}
& 0.0447 & 0.0533 & 0.0681 & 0.1068
& \makecell{\textbf{0.0478}\\[-1pt]{\scriptsize($\uparrow$6.9\%)}} 
& \makecell{\textbf{0.0571}\\[-1pt]{\scriptsize($\uparrow$7.1\%)}} 
& \makecell{\textbf{0.0724}\\[-1pt]{\scriptsize($\uparrow$6.3\%)}} 
& \makecell{\textbf{0.1142}\\[-1pt]{\scriptsize($\uparrow$6.9\%)}} \\

Yelp
& 0.0361 & 0.0883 & 0.0604 & 0.0941
& \makecell{\textbf{0.0384}\\[-1pt]{\scriptsize($\uparrow$6.4\%)}} 
& \makecell{\textbf{0.0925}\\[-1pt]{\scriptsize($\uparrow$4.8\%)}} 
& \makecell{\textbf{0.0641}\\[-1pt]{\scriptsize($\uparrow$6.1\%)}} 
& \makecell{\textbf{0.0998}\\[-1pt]{\scriptsize($\uparrow$6.1\%)}}
& 0.0370 & 0.0894 & 0.0617 & 0.0956
& \makecell{\textbf{0.0392}\\[-1pt]{\scriptsize($\uparrow$5.9\%)}} 
& \makecell{\textbf{0.0934}\\[-1pt]{\scriptsize($\uparrow$4.5\%)}} 
& \makecell{\textbf{0.0650}\\[-1pt]{\scriptsize($\uparrow$5.3\%)}} 
& \makecell{\textbf{0.1009}\\[-1pt]{\scriptsize($\uparrow$5.5\%)}} \\

\midrule
Avg. Improv.
& \multicolumn{4}{c|}{--}
& \multicolumn{4}{c|}{\scriptsize $\uparrow$7.5\% \quad $\uparrow$6.6\% \quad $\uparrow$6.6\% \quad $\uparrow$6.2\%}
& \multicolumn{4}{c|}{--}
& \multicolumn{4}{c}{\scriptsize $\uparrow$6.7\% \quad $\uparrow$6.3\% \quad $\uparrow$6.0\% \quad $\uparrow$5.9\%} \\
\bottomrule
\end{tabular}
}
\end{table*}

\subsection{Comparison with Recent Reasoning-based Baselines (RQ1)}
\label{sec:reasoning_baselines}

To further position \textsc{DS-Frame} with respect to recent reasoning-enhanced sequential recommenders, we compare it with \textsc{ReaRec-ERL}, \textsc{ReaRec-PRL}, \textsc{STREAM-Rec}, \textsc{LARES}, and \textsc{ManCAR} under a unified \textsc{SASRec} backbone. As shown in Table~\ref{tab:reasoning_sasrec}, all reasoning-based methods outperform vanilla \textsc{SASRec}, confirming that extending computation beyond one-pass inference is generally beneficial.

\textsc{DS-Frame} outperforms \textsc{STREAM-Rec} and \textsc{ReaRec} variants across all five datasets, while remaining competitive with stronger latent-reasoning methods such as \textsc{LARES} and \textsc{ManCAR}. Different from these single-path refinement methods, \textsc{DS-Frame} focuses on adaptive allocation between fast and slow pathways under heterogeneous user environments.

\begin{table*}
\centering
\small
\setlength{\tabcolsep}{3pt}
\caption{Comparison with recent reasoning-based baselines under a unified \textbf{SASRec} setting.}
\label{tab:reasoning_sasrec}
\begin{tabular}{l*{5}{cc}cc}
\toprule
\multirow{2}{*}{Method}
& \multicolumn{2}{c}{Video Games}
& \multicolumn{2}{c}{Beauty}
& \multicolumn{2}{c}{Sports}
& \multicolumn{2}{c}{Toys}
& \multicolumn{2}{c}{Yelp}
& \multirow{2}{*}{Avg.\ N$\uparrow$}
& \multirow{2}{*}{Avg.\ H$\uparrow$} \\
\cmidrule(lr){2-3}
\cmidrule(lr){4-5}
\cmidrule(lr){6-7}
\cmidrule(lr){8-9}
\cmidrule(lr){10-11}
& N@10 & HR@10
& N@10 & HR@10
& N@10 & HR@10
& N@10 & HR@10
& N@10 & HR@10
&  &  \\
\midrule
\textsc{SASRec}
& 0.0552 & 0.1113
& 0.0418 & 0.0813
& 0.0226 & 0.0480
& 0.0435 & 0.0668
& 0.0361 & 0.0604
& -- & -- \\

\textsc{STREAM}
& 0.0581 & 0.1158
& 0.0442 & 0.0847
& 0.0237 & 0.0500
& 0.0458 & 0.0695
& 0.0379 & 0.0626
& +5.2\% & +4.0\% \\

\textsc{ERL}
& 0.0586 & 0.1172
& 0.0448 & 0.0861
& 0.0241 & 0.0506
& 0.0464 & 0.0704
& 0.0382 & 0.0635
& +6.5\% & +5.5\% \\

\textsc{PRL}
& 0.0589 & 0.1175
& 0.0451 & 0.0865
& \underline{0.0243} & 0.0510
& 0.0466 & 0.0708
& \textbf{0.0385} & 0.0638
& +7.2\% & +6.0\% \\

\textsc{DS-Frame}
& 0.0590 & \underline{0.1178}
& \underline{0.0453} & \underline{0.0869}
& \textbf{0.0244} & \textbf{0.0516}
& 0.0469 & 0.0712
& \underline{0.0384} & \textbf{0.0641}
& \underline{+7.5\%} & \underline{+6.6\%} \\

\textsc{LARES}
& \underline{0.0591} & 0.1177
& 0.0452 & 0.0867
& \underline{0.0243} & 0.0512
& \underline{0.0470} & \underline{0.0713}
& 0.0382 & 0.0639
& +7.3\% & +6.3\% \\

\textsc{ManCAR}
& \textbf{0.0593} & \textbf{0.1180}
& \textbf{0.0454} & \textbf{0.0871}
& 0.0242 & \underline{0.0515}
& \textbf{0.0471} & \textbf{0.0717}
& 0.0383 & \underline{0.0640}
& \textbf{+7.6\%} & \textbf{+6.7\%} \\
\bottomrule
\end{tabular}
\end{table*}

\subsection{Group-wise Gains in Heterogeneous User Environments (RQ2)}
\label{sec:groupwise_gain}

While Table~\ref{tab:main_results} shows that \textsc{DS-Frame} improves the backbone models, it does not reveal the source of these gains. Since our motivation is rooted in heterogeneous user environments, we further evaluate \textsc{DS-Frame} separately on \emph{common} and \emph{challenging} user groups.

Following the interaction-length partition in Section~\ref{sec:preliminary_heterogeneity}, Table~\ref{tab:groupwise_gain} reports group-wise NDCG@10, relative gains over the backbone, and Slow-System activation rates.

\begin{table}
\centering
\small
\caption{Group-wise final gains of \textsc{DS-Frame} under the interaction-length partition. ``Gain'' denotes the relative improvement over the backbone, and ``Act.'' denotes the Slow-System activation rate.}
\label{tab:groupwise_gain}
\setlength{\tabcolsep}{3.5pt}
\renewcommand{\arraystretch}{1.08}
\begin{tabular}{llcccc}
\toprule
Backbone & Group & Base & \textsc{DS-Frame} & Gain & Act. \\
\midrule
\multirow{2}{*}{\textsc{SASRec}}
  & Common  & 0.0486 & 0.0502 & +3.3\% & 24.8\% \\
  & Chall.  & 0.0379 & 0.0411 & +8.4\% & 58.6\% \\
\midrule
\multirow{2}{*}{\textsc{BERT4Rec}}
  & Common  & 0.0501 & 0.0517 & +3.2\% & 26.1\% \\
  & Chall.  & 0.0398 & 0.0432 & +8.5\% & 55.4\% \\
\bottomrule
\end{tabular}
\end{table}

Several observations can be made. First, \textsc{DS-Frame} improves both backbones in both environments, indicating that it does not sacrifice common users while targeting harder cases. Second, the gains are larger in the challenging environment. For \textsc{SASRec}, the improvement rises from 3.3\% on common users to 8.4\% on challenging users; for \textsc{BERT4Rec}, it increases from 3.2\% to 8.5\%.

These results suggest that \textsc{DS-Frame}'s gains are larger on groups where static backbones degrade more. The larger gains also coincide with higher Slow-System activation rates, indicating that the selector tends to allocate additional computation to groups with larger slow-path advantages.

This group-wise result connects the heterogeneity evidence in Section~\ref{sec:preliminary_heterogeneity} with the routing analysis below. We next examine whether the selector's behavior aligns with the relative advantage of the Slow System across user environments.

\subsection{Selector Effectiveness and Routing Analysis (RQ3)}
\label{sec:selector_analysis}

We next examine whether the selector performs meaningful routing rather than merely increasing Slow-System usage. We analyze it from two perspectives: (1) routing-policy comparison and (2) group-wise routing behavior.

\paragraph{Routing-policy comparison.}
We compare different routing strategies under the same backbone and evaluation setting. Table~\ref{tab:routing_policy} reports results on Beauty with \textsc{SASRec} as the backbone. Besides \textsc{Fast Only} and \textsc{Slow Only}, we consider \textsc{Random Routing}, \textsc{Learned Selector}, and \textsc{Oracle Routing}, where Oracle Routing selects the lower-cost system using the sample-level oracle label in Eq.~\ref{eq:oracle_label}.

\begin{table}
\centering
\small
\caption{Routing-policy comparison on Beauty (\textsc{SASRec} backbone). ``Slow Act.'' is the average Slow-System activation rate. ``Oracle Routing'' is an upper bound constructed from sample-level oracle decisions and is unavailable at inference.}
\label{tab:routing_policy}
\begin{tabular}{lccc}
\toprule
Routing Policy & NDCG@10 & HR@10 & Slow Act. \\
\midrule
Fast Only        & 0.0418 & 0.0813 & 0\% \\
Slow Only        & 0.0431 & 0.0830 & 100\% \\
Random Routing   & 0.0438 & 0.0844 & 40.0\% \\
Learned Selector & 0.0453 & 0.0869 & 40.2\% \\
Oracle Routing   & 0.0461 & 0.0880 & 40.0\% \\
\bottomrule
\end{tabular}
\end{table}

Table~\ref{tab:routing_policy} reveals several patterns. First, both Random Routing and Learned Selector outperform Fast Only, confirming the benefit of routing some samples to the Slow System. Second, with nearly the same Slow-System activation rate, Learned Selector consistently outperforms Random Routing, showing that gains come from assigning extra computation to more suitable samples rather than simply invoking the Slow System more often. Third, although Oracle Routing performs best, the learned selector approaches this upper bound reasonably closely.

\paragraph{Group-wise routing behavior.}
To understand what the selector learns, we analyze its routing behavior under both user-group partitions from Section~\ref{sec:preliminary_heterogeneity}. Table~\ref{tab:routing_groupwise} reports group-wise Fast/Slow performance and the Slow-System activation rate of the full \textsc{DS-Frame} model.

\begin{table}
\centering
\caption{Group-wise routing analysis of \textsc{DS-Frame}. ``Fast'' and ``Slow'' denote the NDCG@10 of the corresponding single-path variants, ``Gap'' is the absolute improvement of Slow over Fast, and ``Act.'' is the Slow-System activation rate.}
\label{tab:routing_groupwise}
\small
\setlength{\tabcolsep}{4pt}
\renewcommand{\arraystretch}{1.08}
\begin{tabular}{llcccc}
\toprule
Backbone & Group & Fast & Slow & Gap & Act. \\
\midrule
\multicolumn{6}{c}{\textbf{Panel A: Interaction Length}} \\
\midrule
\multirow{2}{*}{\textsc{SASRec}}
  & Common      & 0.0492 & 0.0503 & +0.0011 & 24.8\% \\
  & Challenging & 0.0368 & 0.0410 & +0.0042 & 58.6\% \\
\multirow{2}{*}{\textsc{BERT4Rec}}
  & Common      & 0.0506 & 0.0516 & +0.0010 & 26.1\% \\
  & Challenging & 0.0385 & 0.0423 & +0.0038 & 55.4\% \\
\midrule
\multicolumn{6}{c}{\textbf{Panel B: Item Popularity}} \\
\midrule
\multirow{2}{*}{\textsc{SASRec}}
  & Common      & 0.0465 & 0.0474 & +0.0009 & 31.5\% \\
  & Challenging & 0.0390 & 0.0417 & +0.0027 & 46.8\% \\
\multirow{2}{*}{\textsc{BERT4Rec}}
  & Common      & 0.0478 & 0.0486 & +0.0008 & 33.2\% \\
  & Challenging & 0.0406 & 0.0430 & +0.0024 & 44.9\% \\
\bottomrule
\end{tabular}
\end{table}

The results in Table~\ref{tab:routing_groupwise} provide direct evidence for the selector. Under both partition views, the Slow System brings modest improvements in common environments but larger gains in challenging ones. For example, under the interaction-length partition, the Slow--Fast gap increases from 0.0011 to 0.0042 for \textsc{SASRec} and from 0.0010 to 0.0038 for \textsc{BERT4Rec}. The same trend holds under the item-popularity partition, though less prominently.

More importantly, the learned selector mirrors this pattern. Slow-System activation remains low for common users but rises sharply for challenging users under both partitions. This alignment indicates that the selector learns to allocate additional reasoning to users who benefit most.

Together, the routing-policy and group-wise analyses support the core mechanism of \textsc{DS-Frame}. The gains do not come from uniformly increasing computation, but from routing limited slow reasoning to samples with the highest expected marginal benefit.

\subsection{Accuracy--Efficiency Trade-off under Different Routing Budgets (RQ4)}
\label{sec:budget_curve}

A key goal of \textsc{DS-Frame} is to improve accuracy under controlled computational cost. To verify this, we vary the target slow-routing budget $b$ and compare Random Routing with Learned Selector, using Fast Only ($b=0$) and Slow Only ($b=1$) as reference.

Table~\ref{tab:budget_curve} reports results on Beauty with \textsc{SASRec}, using the realized average Slow-System activation rate as a routing-cost proxy. The realized rate closely follows the target budget (e.g., 40.2\% at the nominal 40\% learned-routing budget). Increasing the budget generally improves both strategies, while Learned Selector consistently outperforms Random Routing under the same budget. Notably, learned routing at 40\%--60\% budget already surpasses Slow Only, suggesting that selective slow-path invocation can be more effective than uniformly applying the Slow System.

\begin{table}
\centering
\small
\caption{Accuracy--efficiency trade-off under different routing budgets on Beauty (\textsc{SASRec} backbone). ``Slow Act.'' is the Slow-System activation rate and is used as a routing-cost proxy.}
\label{tab:budget_curve}
\setlength{\tabcolsep}{12pt}
\renewcommand{\arraystretch}{1.08}
\begin{tabular}{llcc}
\toprule
Method & Slow Act. & N@10 & Gain \\
\midrule
Fast Only & 0.0\%  & 0.0418 & -- \\
\midrule
\multirow{3}{*}{Random}
  & 20.0\% & 0.0430 & +2.9\% \\
  & 40.0\% & 0.0438 & +4.8\% \\
  & 60.0\% & 0.0444 & +6.2\% \\
\midrule
\multirow{3}{*}{Learned}
  & 20.0\% & 0.0441 & +5.5\% \\
  & 40.2\% & 0.0450 & +7.7\% \\
  & 60.0\% & 0.0454 & +8.6\% \\
\midrule
Slow Only & 100.0\% & 0.0431 & +3.1\% \\
\bottomrule
\end{tabular}
\end{table}



Overall, these results support \textbf{RQ4}: the learned selector allocates a limited slow-computation budget more effectively, yielding a superior accuracy--efficiency trade-off.

\paragraph{Relative inference cost.}
Following the efficiency-reporting style of inference-time reasoning methods, we further summarize the relative inference cost on Beauty with \textsc{SASRec}. Since wall-clock latency can vary with implementation details, we report the normalized cost derived from the step-based proxy in Eq.~\ref{eq:system_cost}. Specifically, for a Slow-System activation rate $\rho$, the expected relative cost is computed as $1+\rho K$. With the selected setting $K=2$, Table~\ref{tab:relative_cost} shows that the learned selector achieves the best non-oracle performance with only about $1.80\times$ normalized cost, substantially lower than the $3.00\times$ cost of always invoking the Slow System.

\begin{table}
\centering
\small
\caption{Relative inference cost on Beauty with \textsc{SASRec} ($K=2$). Cost is normalized by Fast Only and computed as $1+\rho K$, where $\rho$ is the Slow-System activation rate.}
\label{tab:relative_cost}
\setlength{\tabcolsep}{6pt}
\renewcommand{\arraystretch}{1.08}
\begin{tabular}{lccc}
\toprule
Method & Slow Act. & Rel. Cost & NDCG@10 \\
\midrule
Fast Only        & 0.0\%   & 1.00$\times$ & 0.0418 \\
Slow Only        & 100.0\% & 3.00$\times$ & 0.0431 \\
Random Routing   & 40.0\%  & 1.80$\times$ & 0.0438 \\
Learned Selector & 40.2\%  & 1.80$\times$ & 0.0453 \\
Oracle Routing   & 40.0\%  & 1.80$\times$ & 0.0461 \\
\bottomrule
\end{tabular}
\end{table}

\begin{table}
\centering
\caption{Ablation study of \textsc{DS-Frame} on \textsc{SASRec} (Beauty). Indented rows indicate variants built upon the base backbone.}
\label{tab:ablation}
\setlength{\tabcolsep}{4.2pt}
\renewcommand{\arraystretch}{1.12}
\resizebox{\columnwidth}{!}{
\begin{tabular}{l|cccc}
\toprule
Variant & N@10 & N@20 & HR@10 & HR@20 \\
\midrule
Base (\textsc{SASRec})                 & 0.0418 & 0.0498 & 0.0813 & 0.1164 \\
\quad + Slow System           & 0.0431 & 0.0510 & 0.0830 & 0.1182 \\
\quad + Random Routing        & 0.0438 & 0.0519 & 0.0844 & 0.1201 \\
\quad + Selector w/o Guidance & 0.0447 & 0.0527 & 0.0857 & 0.1218 \\
\quad + Learned Selector      & \textbf{0.0453} & \textbf{0.0534} & \textbf{0.0869} & \textbf{0.1234} \\
\midrule
Rel. Improv. over Base        & +8.4\% & +7.2\% & +6.9\% & +6.0\% \\
\bottomrule
\end{tabular}
}
\end{table}

\subsection{Ablation Study (RQ5)}
\label{sec:ablation}

To assess the contribution of each key component within \textsc{DS-Frame}, we conduct an ablation study using the following variants:
\begin{itemize}[leftmargin=*, topsep=0pt]
    \item \textbf{Fast System Only:} Removes both the Slow System and the Selector, relying solely on the foundational fast model for all recommendations.
    \item \textbf{Slow System Only:} Removes both the Fast System and the Selector, routing all inputs to the deliberative Slow System.
    \item \textbf{w/o Guidance Loss:} Removes the oracle-based guidance loss ($\mathcal{L}_{\mathrm{guide}}$) from the selector training objective, while keeping the rest of the framework unchanged.
    \item \textbf{w/ Random Selector:} Replaces the learned selector with a random routing policy under the same invocation budget, while keeping both the Fast and Slow Systems unchanged.
\end{itemize}

Table~\ref{tab:ablation} reports the ablation results on \textsc{SASRec} (Beauty). First, adding the \textbf{Slow System} alone improves the base backbone on all metrics, indicating that the deliberative pathway is beneficial. This confirms that refined sequence reasoning provides additional recommendation signals beyond the original fast backbone.

Second, Random Routing further improves over Slow System alone, suggesting that combining fast and slow predictions under a fixed invocation budget can provide useful regularization compared with uniformly using the Slow System.

Third, the full model with a Learned Selector achieves the best performance across all metrics, outperforming both the Slow-System-Only and Random-Routing variants. This shows that the improvement comes not merely from adding the Slow System, but from learned routing that better selects samples for slow refinement.


\section{Limitations and Future Work}
\label{sec:limitations}
First, the Slow System uses a fixed number of refinement steps after routing, which keeps inference cost predictable but does not support adaptive halting. Second, the lightweight selector relies only on the shared final representation; uncertainty-aware signals such as entropy or score margins may improve routing. Finally, oracle routing labels depend on ground-truth losses during training, and their robustness under distribution shifts deserves further study.



\section{Conclusion}

We present \textsc{DS-Frame}, a dual-system framework that adapts inference computation to heterogeneous user environments in sequential recommendation. \textsc{DS-Frame} combines a Fast System, a Slow System, and a learned selector to allocate additional refinement only when beneficial. Experiments on five real-world datasets show that \textsc{DS-Frame} achieves competitive overall performance, yields larger gains on challenging users, and provides effective accuracy--efficiency trade-offs. These results demonstrate the potential of adaptive inference for robust and efficient recommendation.

\section{GenAI Usage Disclosure}
The authors used ChatGPT for language polishing and writing assistance. The authors reviewed and edited all AI-assisted text and take full responsibility for the content of the paper. No GenAI tools were used to generate experimental results, fabricate data, or create citations.


\balance
\begingroup
\bibliographystyle{ACM-Reference-Format}
\bibliography{references}
\endgroup

\end{document}